\documentstyle[epsf,12pt,aps]{revtex}
\newcommand{\be}{\begin{equation}}
\newcommand{\ee}{\end{equation}}
\newcommand{\bea}{\begin{eqnarray}}
\newcommand{\eea}{\end{eqnarray}}

\begin{document}

\baselineskip=24pt
\begin{center}
{{\bf {\large\bf Equilibrium and stability properties of a coupled
two-component Bose-Einstein condensate} }}
\end{center}
\begin{center}
{\small\sf
 Chi-Yong Lin$^{a,1}$, E.J.V. de Passos$^{b,2}$, M. S. Hussein$^{b,3}$ Da-Shin
 Lee$^{a,4}$ and A.F.R. de Toledo Piza$^{b,5}$}
\end{center}

\begin{center}
{\small\it $^{a}$Department of Physics, National Dong Hwa
University Hua-Lien, Taiwan, R.O.C.}
\\
{\small\it $^{b}$Instituto de F\'{\i}sica, Universidade de S\~ao
Paulo, CP 66318 CEP 05389-970, S\~ao Paulo, SP, Brazil}
\end{center}


\vskip .3cm \baselineskip=18pt
\begin{center}
{\bf ABSTRACT}
\end{center}
\vskip 0.1cm

The equilibrium and stability properties of a coupled
two-component BEC is studied using a variational method and the
one-dimensional model of Williams and collaborators. The
variational parameters are the population  fraction, translation
and scaling transformation of the condensate densities, assumed to
have a Gaussian shape.
We study the equilibrium and stability properties as a function of
the strength of the laser field and the traps displacement.
We find many branches of equilibrium configurations, with a host
of critical points. In all cases, the signature of the onset of
criticality is the collapse of a normal  mode which is a linear
combination of the out of phase translation
  and an in phase breathing oscillation of  the  condensate  densities.
Our calculations also indicate that we have symmetry breaking
effects when the traps are not displaced.\\


\vskip 1.5cm

\vfill \hspace{\fill}

\noindent\makebox[66mm]{\hrulefill}

\footnotesize
\indent$^{1}$e-mail: lcyong@mail.ndhu.edu.tw \\
\indent$^{2}$e-mail: passos@fma.if.usp.br \\
\indent$^{3}$e-mail: hussein@if.usp.br \\
\indent$^{4}$e-mail: dslee@mail.ndhu.edu.tw\\
\indent$^{5}$e-mail: piza@fma.if.usp.br 
\newpage
\normalsize \baselineskip=20pt

\bigskip

\section {Introduction}

One of the most interesting achievements in the field of
boson condensation was the experimental observation of binary
mixture of trapped condensates [1-4]. The species in the mixture
can be atoms with different F spin orientations [3-4] or simply
different hyperfine states of the same atom [1-2]. In the last case
we can have inter-conversion between the components through the
coupling of the atoms with an external laser field [5].

\indent{In} the case of uncoupled components many effects have
been predicted theoretically and determined experimentally such as
spatial phase separation [6], stability properties of the
equilibrium states [6-7] and symmetry breaking effects [6,8].
These aspects of binary condensate mixtures have been treated in
the literature using various methods such as the Thomas-Fermi (TF)
approximation [8-9] and numerical solutions of the appropriate
coupled Gross-Pitaevskii (GP) equations [10].

\indent{In} this paper we use the variational method [11-13] to
study the equilibrium properties of a coupled two-component BEC.
In our exploratory calculation we use the one-dimensional model of
reference [14] and take as variational parameters the population
fraction, translation and scaling transformation of the
equilibrium state densities, all assumed to have a Gaussian shape.
We consider the intensity of the interaction between the
condensates equal and the detuning is put equal to zero, leading
to equal equilibrium population fraction. In our calculation we
investigate the behavior of the spatial phase separation as a
function of the trap displacement and the strength of the laser field.
The coupling with the laser field have a stabilizer effect in
the process of phase separation, opposite to the one coming from
the trap displacement.Our paper is organized as follows:In section 2
we show briefly how to use the variational method to study the equilibrium
and stability properties of a binary mixture of
condensates.Specifically,we derive the general form of the equations
of motion and develop a scheme to find the normal modes.In section 3
we present our numerical results and discuss its physical
significance.A summary of our numerical results is presented in
section 4.

\section {Variational study of the equilibrium and stability
properties of a binary mixture of condensates.}

\subsection { The variational approach}

\indent{The} starting point of our discussion is the action [15]
\be S \;=\; \int dz dt\; \Bigl[\;\; \sum_{j} i\hbar \psi^{*}_{j}
(z,t) \dot{\psi}_{j} (z,t) - {\cal E} (z,t)\; \Bigr] \ee

\noindent{where} $\psi_{j} (z, t)$, $j=a,b$ are the condensate
wave function of each component in the mixture and ${\cal E}
(z,t)$ is the energy density,
\be {\cal E}(z,t)= \sum_{j}
\psi_j^*\Bigl[-\frac{\hbar^2}{2m}\frac{d^2}{dz^2}
     +V^{j}_{trap}\Bigr]\psi_j
+\frac{1}{2}\sum_{k,j}\lambda_{kj}|\psi_j|^2|\psi_k|^2 +\Omega
\Bigl[\psi^*_a\psi_b+\psi^*_b\psi_a\Bigr] \;. \ee

\noindent{In} the above expression $V^j_{trap} (z)$, $j=a,b$ are
the trapping potential of each component
\be V^j_{trap}=\frac{m}{2}\omega^2_z(z+\gamma_j z_0)^2  \ee

\noindent{where} $\gamma_a=-1$ and $\gamma_b=1$, $z_0$ is the trap
displacement, $\lambda_{aa}=\lambda_{bb}$ and $\lambda_{ab}$ are
the strength of the intraspecies and interspecies interactions,
respectively, and the last term comes from the coupling with the laser
field
responsible by the interspecies tunnelling, with $\Omega$ the
Rabi frequency.

\indent{Imposing} that the action (1) is stationary with respect
to a variation of $\psi_j(z,t)$ subject only to the normalization
constrain $\sum_{j}|\psi_j(z,t)|^2=N$, leads to the coupled
time-dependent Gross-Pitaevskii equations for the condensate wave
functions [15],
\begin{eqnarray}
i\hbar\frac{\partial\psi_a}{\partial
t}&=&\Bigl(-\frac{\hbar^2}{2m}\frac{\partial^2}{\partial
z^2}+V_{trap}^{(a)}
+\lambda_{aa}|\psi_a|^2+\lambda_{ab}|\psi_b|^2\Bigr)\psi_a
+\Omega\psi_b \; \\
i\hbar\frac{\partial\psi_b}{\partial
t}&=&\Bigl(-\frac{\hbar^2}{2m}\frac{\partial^2}{\partial
z^2}+V_{trap}^{(b)}
+\lambda_{ab}|\psi_a|^2+\lambda_{bb}|\psi_b|^2\Bigr)\psi_b
+\Omega\psi_a
\end{eqnarray}

\noindent{To} get the equations for the stationary states, we look for
solutions of equations (4) and (5) of the form 
$\psi_j (z,t)=e^{-i\frac{\mu t}{\hbar}}\psi_j (z)$ ,which gives rise
to the time-independent GP equations:
\begin{eqnarray}
\mu\psi_a&=&\Bigl(-\frac{\hbar^2}{2m}\frac{d^2}{d
z^2}+V_{trap}^{(a)}
+\lambda_{aa}|\psi_a|^2+\lambda_{ab}|\psi_b|^2\Bigr)\psi_a
+\Omega\psi_b \; \\
\mu\psi_b&=&\Bigl(-\frac{\hbar^2}{2m}\frac{d^2}{d
z^2}+V_{trap}^{(b)}
+\lambda_{ab}|\psi_a|^2+\lambda_{bb}|\psi_b|^2\Bigr)\psi_b
+\Omega\psi_a
\end{eqnarray}

\indent{We} can view the two condensate wave functions as
components of a spinor of ``quasi-spin'' equal to $\frac{1}{2}$
[16] which leads immediately to the property that the stationary
equations are invariant under a transformation which is the product
of a space reflection, $P_z$, and an atom exchange, $P_{ex}$,
where $P_{ex}=i\exp-i\frac{\pi}{2}\sigma_x$. Therefore, as
$(P_zP_{ex})^2=P_zP_{ex}$, we can classify the stationary states
as even (gerade) and odd (umgerade) under this transformation. In
the even class $\psi^g_a(z)=\psi^g_b(-z)$, whereas in the odd
class $\psi^u_a(z)=-\psi^u_b(-z)$. The next step is to solve
numerically the coupled stationary equations for the condensate
wave functions to map the solutions as a function of $\Omega$ and
$z_0$. However, before embarking on this task, we found it
worthwhile to adopt a simpler approach, the variational method. In
the variational method the search for stationary states reduces to
finding the stationary points in a finite dimensional energy
surface which is much simpler than the corresponding search in an
infinite dimensional space, when we solve exactly the coupled GP
equations. Besides, the variational solutions can be used as an
initial guess in the numerical solution of the coupled GP
equations [10].

\indent{In} the variational approach we parametrize the time
dependence of the condensate wave function through a set of $2d$
parameters [12-13] which we denote by ${\bf w}=({\rm w}_1,{\rm
w}_2,\ldots, {\rm w}_{2d})$,
\begin{equation}
\psi_j(z,t)=\psi_j(z,{\bf w}(t))
\end{equation}

\noindent{When} we replace the condensate wave functions
parametrized as in equation (8), in equation (1), the action
reduces to a ``classical'' action, whose variation leads to
Hamilton-type equations of motion in terms of these parameters[18]
\begin{equation}
\sum_{l}\Gamma_{kl}({\bf w})\; \dot {\rm w}_l = \; \frac{\partial
E}{\partial {\rm w}_k}({\bf w})\;,
\end{equation}

\noindent{where} $E({\bf w})$ is the spatial integral of the
energy density (2),  with the condensate wave functions
parametrized as in equation (8)
\begin{equation}
E({\bf w})=\int dz {\cal E}(z,{\bf w})
\end{equation}

\noindent{and} the antisymmetric matrix $\Gamma_{kl}({\bf w})$ is
given by
\begin{equation}
\Gamma_{kl}({\bf w})=-2\;{\rm Im}\sum_{j}\int dz
\frac{\partial\psi_j^*}{\partial{\rm w}_k}(z,{\bf
w})\frac{\partial\psi_j}{\partial {\rm w}_l}(z,{\bf w})
\end{equation}

\noindent{It} follows from equation (9) that the equilibrium
configurations are determined by the equations
\begin{equation}
\frac{\partial E}{\partial{\rm w}_k}({\bf w}^0)=0\;,
\hspace{0.8cm} k=1,2,\ldots, 2d \; .
\end{equation}

\subsection { Normal Modes}

\indent{To} investigate the stability of the equilibrium
configurations, we calculate the energies of the normal modes.
They are stable if the energies are real and positive and unstable
if one of the energies is complex. To find the normal modes, we
linearize the equations of motion in the neighborhood of an
equilibrium configuration, leading to:
\begin{equation}
\sum_{l}\Gamma_{kl}\dot{\bar{\rm w}_l}(t) =
\sum_{l}{\rm H}_{kl} \bar{\rm w}_{l}\;.
\end{equation}

\noindent{In} this equation $\bar{\bf w}$ are the displacements
from equilibrium ${\bf w}={\bf w}^{0}+\bar{\bf w}$ and ${\rm
H}_{kl}$, $\Gamma_{kl}$ are, respectively, the Hessian and the
matrix $\bf\Gamma({\bf w})$ evaluated at the equilibrium
configuration, that is
\begin{equation}
{\rm H}_{kl}\equiv{\rm H}_{kl} ( {\bf w}^0)=\frac{\partial^2 E
}{\partial{\rm w}_k\partial{\rm w}_l}( {\bf w}^0)
\end{equation}

\noindent{and}

\begin{equation}
\Gamma_{kl}=\Gamma_{kl}({\bf w}^0)
\end{equation}

\indent{To} proceed, we divide the $2d$ parameters into two
groups,  ${\bf w}=(q_1,q_2,\ldots,q_d,p_1,p_2,\ldots,p_d)=({\bf
q},{\bf p})$, of coordinates $\bf q$ and momenta $\bf p$.
Schematically, the reasoning behind this splitting is that the
amplitude of the condensate wave function, whose square is the
condensate density, depends only on the coordinates while its
phase, whose gradient is the velocity field, depends on the
momenta and the coordinates.

\indent{In} terms of the coordinates and  the momenta the
equations of motion (13), read
\begin{equation}
{\bf\Gamma}\;\left( \begin{array}{l} \dot{\bar{\bf q}} \\
\dot{\bar{\bf p}}
\end{array} \right)
\; \; = \; \;
{\bf{ H}}  \, \left( \begin{array}{l} \bar{\bf q} \\ \bar{\bf p}
\end{array} \right) \; \; \; \; .
\end{equation}
\noindent{where} $\bf\Gamma$ and ${\bf H}$ are, respectively,
antisymmetric and symmetric matrices which can be written in terms
of four $d\times d$ blocks
\begin{equation}
{\bf\Gamma} \; = \; \left( \begin{array}{ll} {\bf\Gamma}_{qq}
\;&\; {\bf\Gamma}_{qp}
\\ {\bf\Gamma}_{pq} \;&\; {\bf\Gamma}_{pp}
\end{array}
\right) \; \;  , \; \; \;
{\bf H} \; = \; \left(
\begin{array}{ll} {\bf H}_{qq} \;&\; {\bf H}_{qp}
\\  {\bf H}_{pq} \;&\; {\bf H}_{pp}
\end{array} \right)
\end{equation}

\noindent{where}, for example, ${\bf\Gamma}_{qq}$ and ${\bf
H}_{qq}$ are $d\times d$ matrices whose elements are given by
equations (11) and (14), where the derivatives are with respect to
the coordinates, with an analogous definition for the other
$d\times d$ matrices.

\indent{In} our method to find the normal modes, we try to stay as
close as possible to the one adopted in the standard case [17]. To
begin with, we should find a transformation to a set of new
coordinates and momenta
\begin{equation}
\left( \begin{array}{l} \bar{\bf q} \\ \bar{\bf p}
\end{array} \right)
\; \; = \; \;
{\bf T}^{-1}  \, \left( \begin{array}{l} {\bf Q} \\ {\bf P}
\end{array} \right) \; \; \; \; .
\end{equation}
\noindent{such} that
\begin{equation}
{\bf \Gamma}^{-1}{\bf H}{\bf T}^{-1} \; = \; {\bf T}^{-1}\left(
\begin{array}{ll} {\bf O} \;&\; {\bf\Lambda}
\\ -{\bf\Lambda} \;&\; {\bf O}
\end{array} \right)
\end{equation}

\noindent{where} ${\bf \Lambda}$ is a diagonal $d\times d$ matrix
whose diagonal elements are the normal mode energies.
%

\indent{In} terms of the new coordinates and momenta, the
equations of motion, (16), reduce to
\begin{equation}
\left( \begin{array}{l} \dot{\bf Q} \\ \dot{\bf P}
\end{array} \right)
\; \; = \; \;
\left(
\begin{array}{ll} {\bf 0} \;&\; {\bf \Lambda}
\\ -{\bf\Lambda} \;&\; {\bf 0}
\end{array} \right)
\, \left( \begin{array}{l} {\bf Q} \\ {\bf P}
\end{array} \right) \; \; \; \; .
\end{equation}

\noindent{leading} to
\begin{equation}
\dot{Q}_{k}=\Lambda_{k}P_{k}\;\;, \dot{P}_{k}=-\Lambda_{k}Q_{k}
\end{equation}

\indent{The} equations (18)-(19) define the transformation to the
normal modes and our scheme to find the matrix ${\bf T}^{-1}$ is a
straightforward generalization of the standard case [17]. The
starting point is to solve the eigenvalue problem
\begin{equation}
{\bf\Xi}^{-1}{\bf H} V^{(k)} \; = \; \Lambda_k V^{(k)}
\end{equation}

\noindent{where} ${\bf \Xi}$ is the hermitian and antisymmetric
matrix ${\bf\Xi}=-i{\bf\Gamma}$

\indent{As} in the standard case, this eigenvalue problem
has two properties: (i) if $V^{(k)}$ is an eigenvector with
eigenvalue $\Lambda_k$, then $V^{(k)*}$ is also an eigenvector
with eigenvalue $-\Lambda_{k}^*$. (ii) The eigenvectors with
different eigenvalues are $\bf\Xi$ orthogonal, that is
$V^{(l)+}{\bf\Xi} V^{(k)}=0$, if $\Lambda_l\neq\Lambda^*_k$.

\indent{If} one of the eigenvalues of equation (22) is complex,
the system is unstable. If all the eigenvalues are real it is
stable and we can find the transformation to the normal modes as
follows. We define a matrix ${\bf S}^{-1}$, whose first $d$
columns are the $2d$ components of the $d$ eigenvectors with real
positive eigenvalues and the next $d$ columns, the $2d$ components
of the corresponding eigenvectors with real negative energies.

\indent{In} terms of ${\bf S}^{-1}$, the eigenvalue equations (22)
reads

\begin{equation}
{\bf \Xi}^{-1}{\bf H}{\bf S}^{-1} \; = \; {\bf S}^{-1}\left(
\begin{array}{ll} {\bf\Lambda} \;&\; {\bf 0}
\\ {\bf 0} \;&\; -{\bf\Lambda}
\end{array} \right)
\end{equation}

\noindent{and} the orthogonality of the eigenvectors gives that
\begin{equation}
{\bf S}^{-1\dag}{\bf\Xi}{\bf S}^{-1} \; = \; \left(
\begin{array}{ll} {\bf 1} \;&\; {\bf 0}
\\ {\bf 0} \;&\; -{\bf 1}
\end{array} \right)
\end{equation}

\noindent{The} matrix ${\bf T}^{-1}$ is related to ${\bf S}^{-1}$
by
\begin{equation}
{\bf T}^{-1}={\bf S}^{-1}{\bf U}
\end{equation}

\noindent{where} ${\bf U}$ is the unitary matrix
\begin{equation}
{\bf U} \; = \; \frac{1}{\sqrt{2}}\left(
\begin{array}{ll} {\bf 1} \;&\; {\bf i}
\\{\bf 1} \;&\; -{\bf i}
\end{array} \right)
\end{equation}

\indent{To} summarize, we state the main steps in our procedure:
(i) First we determine the equilibrium configurations by solving
the $2d$ equations (12). (ii) Next, for each equilibrium
configuration, we solve the eigenvalue equation (22). (iii) When
the configuration is stable, we construct the matrix ${\bf S}^{-1}$
from the eigenvectors as indicated above and ${\bf T}^{-1}$ as
shown in equation (25). When the system oscillates in a normal mode
only one pair $(Q_k,P_k)$ is different from zero and equation (18)
gives the time evolution of the system in this case.

\subsection{ Gaussian ansatz} 

\indent{Now} that we have established the general framework of our
calculations, we turn to our specific application where the
variational parameters are related to the population fraction,
translation and breathing shape oscillation of the condensate
densities. Thus, the condensate wave functions are written as
[18]
\begin{equation}
\psi_{j}(z,{\bf w})=e^{iF_j(z,{\bf w})}A_j(z,{\bf w})
\end{equation}

\noindent{where} the amplitude of the condensate wave function is
parametrized as
\begin{equation}
A_j(z,{\bf w})=\sqrt{Nn_j}
\frac{1}{\sqrt{\pi^{1/2}q_{2j}}}e^{-\frac{(z-q_{1j})^2}{2q_{2j}^2}}
\end{equation}

\noindent{and} the phase as
\begin{equation}
F_j(z,{\bf w})=p_{1j}(z-q_{1j})+
\frac{p_{2j}}{2q_{2j}}(z-q_{1j})^2+\theta_j
\end{equation}

\noindent{The} parameters introduced in the above expression have
the following physical interpretation: $n_j$ is the population
fraction of each condensate,
\begin{equation}
Nn_j=\int |\psi_j(z,{\bf w})|^2dz\;,
\end{equation}

\noindent{$q_{1j}$} is the center of mass of the spatial
distribution of each condensate,
\begin{equation}
Nn_jq_{1j}=\int z|\psi_j(z,{\bf w})|^2dz \;,
\end{equation}

\noindent{and} $q_{2j}/\sqrt{2}$ is the width of the spatial
distribution of each condensate
\begin{equation}
Nn_j\frac{q_{2j}^2}{2}=\int (z-q_{1j})^2|\psi_j(z,{\bf w})|^2dz
\;,
\end{equation}

\indent{The} momenta $p_{1j},p_{2j}$ and $\theta_j$ appear in the
phase of the condensate wave functions. The $p_{1j}$ is the center
of mass moment of each condensate

\begin{equation}
Nn_jp_{1j}=-i\int\psi_j^*(z,{\bf w})\frac{\partial\psi_j}{\partial
z}(z,{\bf w})dz
\end{equation}

\noindent{and} $p_{2j}$ is connected to the expectation value of
the dilatation operator in the center of mass frame of each
condensate
\begin{equation}
\frac{1}{2}Nn_{j}p_{2j}q_{2j}=-\frac{i}{2\hbar}\int\Bigl(
\psi^*_j(z,{\bf w})\frac{\partial\psi_j}{\partial z}(z,{\bf
w})-\frac{\partial\psi^*_j}{\partial z}(z,{\bf w})\psi_j(z,{\bf
w})\Bigr)(z-q_{1j})dz\;,
\end{equation}

\noindent{showing} that $q_{1j},p_{1j}$ is related to the
translational degrees of freedom and $q_{2j},p_{2j}$ to the
breathing shape oscillation.

\indent{Since} $E({\bf w})$ depends only on the phase difference 
 between the two condensates, the equations of motion (9) give
that the total number of particles is conserved,
$n_a(t)+n_b(t)=1$. This reduces the number of degrees of freedom
to ten, which we take as the relative phase
$\theta=\theta_b-\theta_a$ and the relative population fraction
$n=\frac{n_b-n_a}{2}$, in addition to the
$q_{1j},q_{2j},p_{1j},p_{2j},j=a,b$.

\section{ Numerical Results}

\indent{To} perform the calculations we specify the model
parameters . Following  reference [14], we
consider our system to be $^{87}$Rb and take $N=2.3\times 10^4$,
$\nu_z=60$Hz which gives $z_{sho}=\sqrt{\hbar/m\omega_z}=1.4 \mu
$m and we put all the interaction strength equal to
$\lambda_{aa}=\lambda_{bb}=\lambda_{ab}=17.5$ $\mu$m sec$^{-1}\hbar$, in order
to reproduce qualitatively the spatial distribution of the condensates
 shown in FIG.2   of reference [14].

\indent{Our} first task is to solve the ten equilibrium equations
(12). Six of the equations lead immediately to
$p^0_{1j}=p^0_{2j}=0,j=a,b$, equal equilibrium population fraction
$n^0=0$, and $\theta^0=0$ or $\pi$. The solution with $\theta^0=0$
belongs to the even class under $P_{z} P_{ex}$ and the ones with
$\theta^0=\pi$, to the odd class. We restrict the calculations to
the odd class since, as $\Omega>0$, the lowest energy
configuration necessarily belongs to this class. Besides, to be an
eigenstate of $P_zP_{ex}$ the equilibrium parameters should obey
the conditions $q_{1a}^0=-q_{1b}^0$ and $q_{2a}^0=q_{2b}^0$.
Therefore, to find the equilibrium configurations we calculate the
zeros of four functions $\frac{\partial E}{\partial q_{kj}}({\rm
w}^0)=0$ with $k=1,2$ and $j=a,b$ and the parameters restricted as
indicated in the above discussion.

\indent{We} characterize the equilibrium configurations (eqc) by
the localization of the center of mass of each component and in
Fig.1 we have a graph of the relative distance between the centers
of mass as a function of $\Omega$, for appropriately chosen values
of $z_0$. As shown in Fig.1, for $z_0=0.23z_{sho}$ we have only
one branch of eqc. For small values of $\Omega/\hbar\omega_z$ the
condensates are well separated and when the intensity of the laser
field increases the overlap between the condensates increases very
slowly up to a value of $\Omega$ when there is a sharp transition
to a mixed phase.

\indent{When} we diminish the value of $z_0$, Fig.1, the behavior
of eqc changes qualitatively. When $\Omega/\hbar\omega_z$ is small
the system behaves as in the previous case. However, when
$\Omega/\hbar\omega_z$ increases there is a critical value of
$\Omega$, $\Omega_{c<}$, where  two branches of eqc appear, one
stable, the other unstable. The two stable eqc  distinguished
by the separation of the centers of mass  are called, respectively,
{\em distant} and {\em near stable} eqc. When we further increase
the value of $\Omega$, the relative distance of the centers of mass
in the unstable eqc increases and  merges with the stable
distant eqc at a critical value of $\Omega$, $\Omega_{c>}$, such
that at $\Omega>\Omega_{c>}$ one is left with only the near stable
eqc, where the condensates are mixed. For smaller values of $z_0$
we have the same pattern, the values of $\Omega_{c<}$ where we
have three branches of eqc and of $\Omega_{c>}$ where we have the
merger of the unstable and distant stable eqc diminishing, this
effect being less pronounced for the latter.

\indent{Also} shown in Fig.1 is the graph of the branches of eqc
for $z_0=0$. We see that already at $\Omega$ near zero we have
three branches of eqc, where now in the near stable eqc there is
complete overlap between the condensates
($\psi_{oa}(z)=\psi_{ob}(z)$). When $\Omega$ increases, the
totally mixed eqc remains, with a density profile independent of
$\Omega$, whereas the distant stable and the unstable eqc approach
each other and merge at $\Omega_{c>}$, such that, at $\Omega >
\Omega_{c>}$ one is left with only the totally mixed eqc. Our
results also show that we have spontaneous symmetry breaking
effects at $z_0=0$ [6,8]. Indeed, at $z_0=0$, our equations are
separately invariant by space reflection, $P_z$, and atom exchange
$P_{ex}$. The totally mixed eqc obey these symmetries separately,
whereas the distant stable eqc do not, being invariant only by the
product of these transformations.

\indent{To} find the signature of the onset of criticality, we
calculate the normal modes along the branches of eqc. We can group
the five normal modes into two sets. In one set there are two
normal modes which are a linear combination of an out-of-phase
oscillation of the centers of mass of each condensate and an
in-phase breathing oscillation  of the condensate densities with
the center of mass of the mixture and the population fraction at
its equilibrium values. In the second set we have three normal
modes which are a linear combination of an in-phase oscillation of
the centers of mass, out-of-phase breathing oscillation of the
condensate densities and particle exchange between the
condensates. The splitting of the normal modes into these two
groups is a general result since it follows from the invariance of
the equations (4-5) under $P_zP_{ex}$, the normal modes of the first
group being even under this transformation and the one of the
second, odd.

\indent{We} found that the signature of criticality in all cases
is the collapse of a normal mode which is an out-of-phase
oscillation of center of mass (dipole oscillation) of the
condensates and an in-phase breathing oscillation of the
condensate densities. In fig.2 we have a graph of the energies of
the two normal modes of the first group, which involves the dipole
oscillation of the condensates, for $z_0=0.08z_{sho}$, which is a
case where we found the existence of critical points (see Fig.1).

\indent{In} Fig.2a and Fig.2b we present a graph of the energies
of the two normal modes along the distant and near stable eqc.
In the distant eqc we
see that there is one normal mode with an almost constant energy, Fig.2a,
and  other whose energy increases beginning from $\Omega=0$,
reaches a maximum value and starts to decrease,Fig.2b. When $\Omega$
approaches the critical value $\Omega_{c>}$, where the distant
stable eqc disappears, the energies of the two normal modes
approaches each other and at $\Omega=\Omega_{c>}$ one of the
energies goes very abruptly to zero. We have a similar behavior in
the graph  of the normal mode energies along the near stable eqc,
Figs.2a-2b. Approaching from above the point where the near stable
eqc disappears, the two energies approaches each other and again
one the of them goes abruptly to zero at $\Omega=\Omega_{c<}$.
This behavior is completely general near a critical point.

\indent{A} ``scar'' of this critical behavior is also present for
a value of $z_0$ at the interface between values where we have and
we do not have critical points, such as $z_0=0.23z_{sho}$ (see
Fig.1). We see in Figs.2c-2d that, corresponding to the very narrow range of
values where the eqc change from separated to mixed, we have also
an abrupt change in the values of the two normal modes energies.
In Fig.3a we detached the region of the sharp change and we see
that it occurs in a very narrow range of values of $\Omega$ and it
is a consequence of a strong level repulsion between the two
approaching even normal mode energies.

\indent{In} Fig.3b we illustrate a generic phenomenon that occurs
when $\Omega\rightarrow 0$, the appearance of a Goldstone zero
energy mode. Indeed, when $\Omega\rightarrow 0$, the particle
number of each component of the mixture is a conserved quantity
and since our theory conserves only the total number of atoms,
this violation is translated into the appearance of a zero energy
mode. Fig.3b shows how the energy of one of odd normal mode goes to
zero for $z_0=0.23z_{sho}$.

\indent{One} question left untouched up to now is the
identification of the lowest energy configuration when we have
many branches. In our model the answer is that, for small
$\Omega$, the distant eqc is always the lowest energy
configuration, changing to the near eqc at a higher value of
$\Omega$, smaller than $\Omega_{c>}$. However, the energy
differences are very small, the equilibrium configurations are
almost degenerate.

\indent{Our} conclusions are based in calculations where we took
$\lambda_{aa}=\lambda_{bb}=\lambda_{ab}$. It is well known that
for homogeneous condensate mixtures, the parameters that controls
the mechanism of phase separation is
$\lambda_{aa}\lambda_{bb}-\lambda_{ab}^2$ [6]. In coupled
mixtures, we add two additional factors which have opposite effects
in the mechanism of spatial separation, the trap displacement and
the laser coupling field and a point that deserves investigation is
how robust are our conclusions when we relax the equal interaction
strength condition.

\section{Summary}

\indent{To} summarize, we have studied the equilibrium and
stability properties of a coupled two-component BEC, as function
of the laser field strength and trap displacement, using the
variational method and the one-dimensional model of reference
[14]. The laser field has a stabilizer effect in the
mechanism of spatial separation of components in the mixture,
opposite to the effect of the trap displacement. We found many
branches of eqc, with a host of critical points. In all cases the
signature of the onset of criticality is the collapse of a normal
mode, which is a linear combination of an out-of-phase translation
and an in-phase breathing oscillation of the condensate densities.

\indent{When} the traps are not displaced, we found eqc which
exhibits symmetry breaking effects. In principle these eqc with
broken symmetry can be reached by, starting at a sufficiently high
value of $\Omega$ and $z_{0}$, adiabatically take the limit $\Omega\rightarrow
0$ and $z_{0}\rightarrow 0$. Taking the limit in the opposite
order, we end up in the symmetric eqc (see Fig.1).

\indent{Undoubtedly} our calculations are  simple. However, it
unveils a very rich structure in systems of coupled condensates,
which should be explored experimentally and theoretically by more
complete calculations.

\bigskip

ACKNOWLEDGMENT: This work was partially supported by
Fundac\c{c}\~ao de Amparo $\grave{a}$ Pesquisa do Estado de
S$\tilde{a}$o Paulo (FAPESP) under contract number 00/06649-9.
EJVP and MSH are supported in part by CNPq.
The work of Chi-Yong Lin and Da-Shin Lee was supported in part by
the National Science Council, ROC under the Grant
NSC-90-2112-M-259-010-Y.EJVP,MSH and AFRTP are members of the
INFOQUANTICA contract 62.0057/01-7-PADCTIII/MIL.


\newpage
\baselineskip=18pt

\begin{figure}[t]
\epsfysize=12cm
\centerline{\epsffile{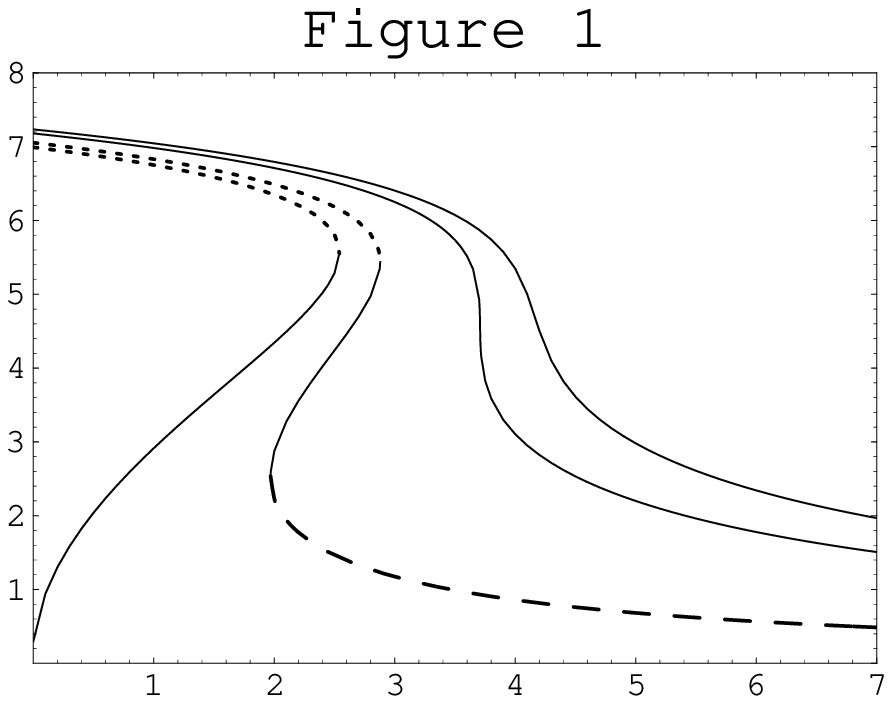}}

\bigskip
\baselineskip=15pt {\sf\noindent{Figure} 1. The plot shows the
relative distance of the centers of mass of the two condensates in
the eqc, $2q_{1a}$, as function of the laser strength
$\Omega$, for fixed values of $z_{0}$. Curves from the top
correspond to $z_0=0.3,0.23,0.08,0 \; z_{sho}$. The dotted and
dashed curves indicate, respectively, the distant and near stable
eqc. For $z_0=0$ the straight line $2q_{1a}=0$ correspond to one
branch of eqc. The laser strength is expressed in units of
$\hbar \omega_z$ and the distance in units of $z_{sho}$. See text
for more details. }

\end{figure}

\newpage
\bigskip

\begin{figure}[t]
\epsfysize=14cm
\centerline{\epsffile{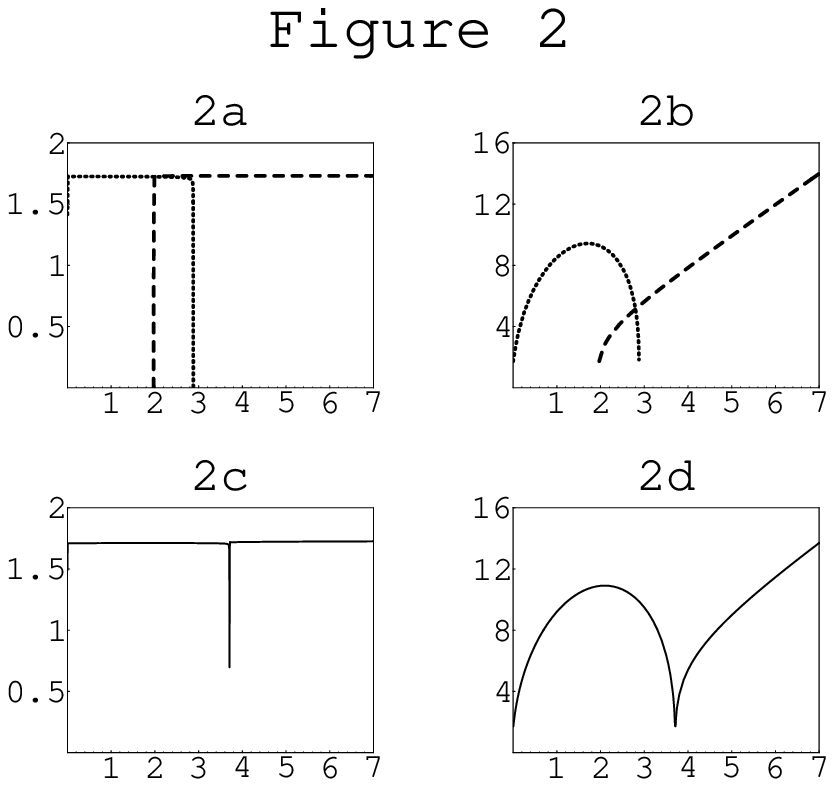}}

\bigskip
\baselineskip=15pt {\sf\noindent{Figure} 2. Fig.2a and Fig.2b show
the energies of the two normal modes, which are an out-of-phase
translation and an in-phase breathing oscillation of the
condensate densities. The energies are calculated along the
$z_0=0.08z_{sho}$ curve, the dotted and the dashed lines
correspond, respectively, to energies along the distant and near
stable eqc. In Figs.2c and 2d., we have a similar graph, now along
the $z_0=0.23z_{sho}$ curve. The energies are measured in units of
$\hbar\omega_z$. See text for more details.}
\end{figure}

\newpage
\bigskip

\begin{figure}[t]
\epsfysize=11cm
\centerline{\epsffile{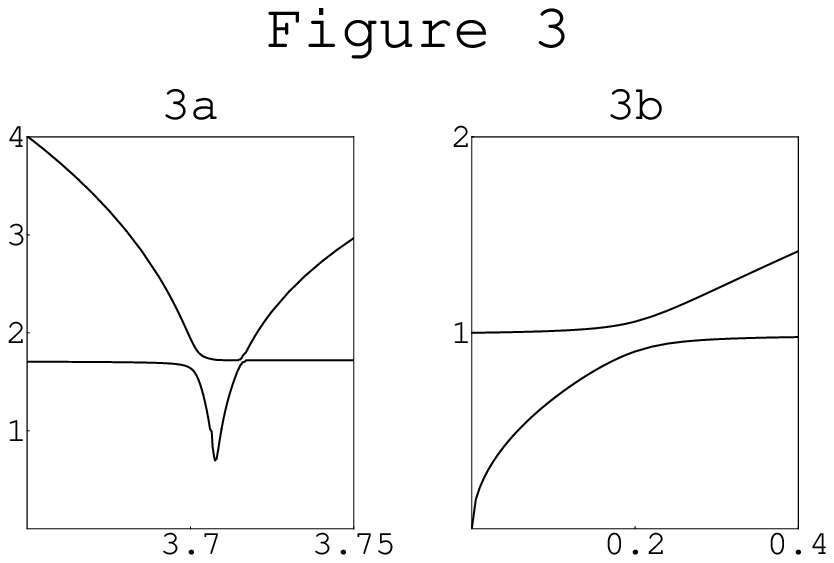}}

\bigskip
\baselineskip=15pt {\sf\noindent{Figure} 3. In Fig.3a we have a
plot of the energies near the point of sharp change, shown in
Figs.2c and 2d. In Fig.3b we have a plot which shows how the
energy of one of the odd normal modes goes to zero when
$\Omega\rightarrow 0$, for $z_0=0.23z_{zsho}$. See text for more
details.}
\end{figure}

\end{document}